\documentclass[epj]{svjour}
\usepackage{graphicx}

\begin{document}

\title{Natural NMSSM with a Light Singlet Higgs and Singlino LSP}
\author{C.T. Potter}
\institute{Physics Department, University of Oregon}
\date{\today}

\abstract{Supersymmetry (SUSY) is an attractive extension of the Standard Model (SM) of particle physics which solves the SM hierarchy problem. Motivated by the theoretical $\mu$-term problem of the Minimal Supersymmetric Model (MSSM), the Next-to MSSM (NMSSM) can also account for experimental deviations from the SM like the anomalous muon magnetic moment and the dark matter relic density. Natural SUSY, motivated by naturalness considerations, exhibits small fine tuning and a characteristic phenomenology with light higgsinos, stops and gluinos. We describe a scan in NMSSM parameter space motivated by Natural SUSY and guided by the phenomenology of an NMSSM with a slightly broken Peccei-Quinn symmetry and a lightly coupled singlet. We identify a scenario which survives experimental constraints with a light singlet Higgs and a singlino lightest SUSY particle. We then discuss how the scenario is not presently excluded by searches at the Large Hadron Collider (LHC) and which channels are promising for discovery at the LHC and International Linear Collider.}


\maketitle

\section{Introduction}

With the discovery of the 125 GeV Higgs boson $h_{125}$ by ATLAS \cite{Aad:2012tfa} and CMS \cite{Chatrchyan:2012ufa} at the Large Hadron Collider (LHC), particle physics enters a new era. In the  Standard Model (SM) of particle physics, the properties of the Higgs boson are determined by theory once the mass is known \cite{Carena:2002es}. At present, their measurements are consistent with the SM prediction \cite{Aad:2015zhl,Khachatryan:2014jba,Aad:2015gba,Aad:2015mxa,Khachatryan:2014kca,Khachatryan:2015mma}.

But the SM is not complete. Experimentally, it does not account for Dark Matter (DM), the anomalous muon magnetic moment or the strong CP problem, among other things. Theoretically, it suffers from the hierarchy problem. Supersymmetry (SUSY) solves the hierarchy problem by introducing a fermionic partner for each SM boson and a bosonic partner for each SM fermion \cite{Martin:1997ns}. 
SUSY with conserved $R$ parity provides a natural candidate for DM, the Lightest Supersymmetric Partner (LSP), and can account for the anomalous muon magnetic moment by introducing new particles in loops. 

The principle of \textit{naturalness} in physics maintains that an effective physical theory approximately valid below some characteristic scale should not be very sensitive to the correct theory above that scale \cite{Dine:2015xga}. Applied to electroweak symmetry breaking in SUSY, this implies that the success of the effective SM Higgs theory disallows SUSY too far above the electroweak scale \cite{Papucci:2011wy}. In particular, the characteristic mass spectrum of \textit{Natural} SUSY includes light superpartners of the Higgs bosons, top quark and gluon near the electroweak scale.

The Minimal SUSY Model (MSSM) contains only the SM particles and their superpartners, together with an enlarged Higgs sector: one neutral pseudoscalar, two neutral scalars and two charged scalars which arise from the two Higgs doublets $\hat{H}_{u}$ and $\hat{H}_{d}$ necessary for the Higgs mechanism in SUSY \cite{Carena:2002es}. But the MSSM suffers from the so-called $\mu$-term problem, which prevents the term  $\mu \hat{H}_{u} \hat{H}_{d}$ in the MSSM superpotential from reaching the electroweak scale without fine tuning \cite{Martin:1997ns,Maniatis:2009re,Ellwanger:2009dp}.

The Next-to MSSM (NMSSM) solves the $\mu$-term problem by introducing a singlet $\hat{S}$ and replacing $\mu \hat{H}_{u} \hat{H}_{d}$ with $\lambda \hat{S} \hat{H}_u \hat{H}_d$. The $Z_{3}$ invariant NMSSM superpotential is \cite{Maniatis:2009re,Ellwanger:2009dp}

\begin{eqnarray}
W &=&\lambda \hat{S} \hat{H}_u\hat{H}_d + \frac{\kappa}{3} \hat{S}^3 
\end{eqnarray}

\noindent where $\lambda$ and $\kappa$ are free parameters. An effective $\mu$-term is generated as the vacuum expectation value of $\hat{S}$, $\mu_{eff}=\lambda \langle \hat{S} \rangle$, reaching a natural scale without fine tuning \cite{Ellwanger:2009dp,Maniatis:2009re}. 

In addition to the Higgs content of the MSSM, the NMSSM contains an additional pseudoscalar and an additional  scalar  so that the NMSSM Higgs sector consists of two neutral pseudoscalars ($a_1,a_2$), three neutral scalars ($h_1,h_2,h_3$) and two charged scalars ($H^+, H^-$) \cite{Miller20043,Ellwanger:2011sk}. The NMSSM Higgs sector is fully determined at tree level by $\lambda$ and $\kappa$, $A_{\lambda}$ and $A_{\kappa}$ (soft trilinear couplings), $\mu_{eff}$  and $\tan \beta$ (ratio of $H_{u},H_{d}$ vacuum expectation values) \cite{Miller20043}. 

One notable version of the NMSSM is the Peccei-Quinn (PQ) symmetric NMSSM, characterized by $\kappa=0$ \cite{Miller:2003hm,Miller20043,Maniatis:2009re,Ellwanger:2009dp}. The PQ symmetric NMSSM explains why there is so little CP violation in the strong sector by exhibiting an axion, the massless pseudoscalar $a_{1}$. In the NMSSM with a slightly broken PQ symmetry, with small $\kappa$ and $A_{\kappa}$, the $a_1$ acquires a small mass proportional to $\kappa A_{\kappa}$  but can still solve the strong CP problem \cite{Miller20043,Hall:2004qd,Schuster:2005py,Barbieri:2007tu,Choi:2013lda}. 

Scenarios with a light NMSSM pseudoscalar Higgs, motivated variously by the strong CP problem, naturalness, the anomalous muon magnetic moment, the $\eta_{b}$ mass spectrum, and the similarity of the baryon density to the dark matter density, have been discussed in the literature \cite{Miller20043,Dermisek:2005ar,Dermisek:2006wr,Dermisek:2007yt,Gunion:2008dg,Domingo:2009tb,Dermisek:2010mg,MarchRussell:2012hi}. 
In this study we assume a light NMSSM pseudoscalar $a_1$ with $2m_{\tau} < m_{a_1} < 2m_{B}$. Motivated by the LEP $Zb\bar{b}$ feature near $m_{b\bar{b}} \approx 60$~GeV \cite{Schael:2006cr}, we identify this as an $h_{1}$ candidate. We further identify the $h_{125}$ as the second lightest neutral scalar $h_2$ of the NMSSM and note that the $h_{125}$ signal strength measurements at the LHC \cite{Aad:2015gba,Khachatryan:2014jba} place the heavier NMSSM $a_2,h_3,H^+$ in the effective MSSM decoupling limit.

\section{Effective MSSM ($\mathbf{\lambda,\kappa \approx 0}$)}

We now consider the phenomenology of the NMSSM with a slightly broken PQ symmetry in which the singlet $S$ is completely decoupled from the doublets $H_u$ and $H_d$ ($\lambda=0$). We then consider how the phenomenology is altered when the singlet is allowed a weak coupling to the doublets ($\lambda \approx 0$). The case $\lambda,\kappa \approx 0$ is known as the effective MSSM \cite{Ellwanger:2009dp}.

For $\lambda=0$, there is no mixing of the singlet with the doublets. The generic couplings in the NMSSM have been detailed in \cite{Franke:1995tc,Ellwanger:2009dp}. We adopt the notation of the latter, denoting $S_{ij}^2$ ($P_{ij}^2$) as the $j$th component of mass eigenstate $h_{i}$ ($a_i$), where $j=1,2,3$ corresponds to the $u$ doublet, the $d$ doublet, and singlet respectively. For purely singlet $h_1$ and $a_1$, $S_{13}=P_{13}=1$ and all other $S_{1j},P_{1j}$ vanish, so the $a_1$ cannot decay to SM particles since their coupling is proportional to $P_{11}=0$ or $P_{12}=0$, and similarly for the $h_1$. The $a_1$ is stable and the only allowed $h_1$ decay for $m_{h_1} \approx 60$~GeV is $h_1 \rightarrow a_1 a_1$. The singlet sector is decoupled from the SM sector.

Furthermore, for the case $\lambda=0$, the singlet sector is decoupled from the MSSM sector. One neutralino is pure singlino whose mass, at tree level, is related to the $a_1$ mass by $m_{\chi} = -2 m_{a_1}^{2}/3A_{\kappa}$ \cite{Miller:2003hm,Miller20043}. For $m_{a_1} \approx 10$~GeV and $\vert A_{\kappa} \vert$ of $\mathcal{O}(1)$~GeV, consistent with a slightly broken PQ symmetry, this yields $m_{\chi} \approx 60$~GeV. In this study we identify the singlino as the LSP $\chi_1$. Denoting $N_{ij}^2$ as the $j$th component of $\chi_i$, where $j=1,2,3,4,5$ corresponds to bino, wino, $u$ higgsino, $d$ higgsino, and singlino respectively. Neutralinos heavier than the singlino LSP have zero singlino component, $N_{15}=1$ and all other $N_{i5}$ vanish. No heavier neutralino can decay to the singlino since the coupling is proportional to $N_{i5}=0$ for $i>1$. The NLSP $\chi_2$ is stable for conserved R parity.  

However, when the singlet is allowed a weak coupling to the doublets ($\lambda \approx 0$), small mixing between the singlet sector and the SM and MSSM sectors is possible. In this case the $a_1$ may couple to SM pairs so the $a_1$ is no longer stable. For $m_{a_1} \approx10$~GeV, $a_1 \rightarrow \tau^+ \tau^-$ dominates, with decays to gluon and light quark pairs subdominant. For $m_{h_1} \approx 60$~GeV, $h_{1} \rightarrow a_1 a_1$ remains dominant with decays to SM pairs, notably $h_1 \rightarrow b\bar{b}$, subdominant. Furthermore, the $\chi_1$ can be produced from heavier neutralino decays since singlino mixing with doublinos is allowed. Then the NLSP $\chi_2$ and NNLSP $\chi_3$ are  no longer stable against decay to $\chi_1$. Above the threshold $m_{\chi_2}=m_{\chi_1}+m_{a_1} \approx 70$~GeV, $\chi_2 \rightarrow \chi_1 a_1$ dominates while below it $\chi_2 \rightarrow \chi_1 Z^{\star}$ dominates. In the latter case, the decay may occur outside the LHC detector effective tracking volume for $\lambda$ of $\mathcal{O}(10^{-2})$ or less \cite{Ellwanger:1997jj}. Above the threshold $m_{\chi_3} =m_{\chi_1}+m_{h_1} \approx 120$~GeV, $\chi_3 \rightarrow \chi_1 h_1$ dominates while below it $\chi_3 \rightarrow \chi_1 a_1$ and $\chi_3 \rightarrow \chi_1 Z^{\star}$ dominate.

Further information about $\lambda$ and $\kappa$ can be extracted from the $h_{1,2}$ sum rule \cite{Miller:2003hm}

\begin{eqnarray}
m_{h_1}^2+m_{h_2}^2 & \approx & m_{Z}^2 + \frac{1}{2} \kappa v_s (4 \kappa v_s + \sqrt{2} A_{\kappa})
\end{eqnarray}

\noindent where $v_{s}\equiv \sqrt{2} \mu_{eff}/\lambda$. For $m_{h_1}=60$~GeV, $m_{h_2}=125$~GeV and $\mu_{eff}=300$~GeV, the sum rule yields $\vert \kappa/\lambda \vert \approx 0.176$.

To summarize, we assume an effective MSSM with mostly singlino LSP $\chi_1$ and $m_{\chi_1} \approx60$~GeV. The $a_1$ and $h_1$ are mostly singlet with dominant decays to SM $\tau$ pairs and $a_1$ pairs, respectively. The $a_1$, $h_1$ and $\chi_1$ can be produced in neutralino decays. For $m_{\chi_2} \approx 70$~GeV or below and $\lambda < \mathcal{O}(10^{-2})$, the $\chi_2$ decays outside of the effective tracking volume. For $m_{\chi_3}\approx 120$~GeV or above, $\chi_3 \rightarrow \chi_1 h_1$ is dominant. Finally, the $h_{1,2}$ mass sum rule yields $\vert \kappa/\lambda \vert \approx 0.176$ for $\mu_{eff}=300$~GeV. These considerations, together with naturalness, inform the parameter ranges in the scan described in the next section.

\section{Parameter Scan}

The parameter scan is performed with NMSSMTools4.4.0 \cite{Ellwanger:2004xm,Ellwanger:2005dv,Belanger:2005kh,Ellwanger:2006rn,Das:2011dg,Muhlleitner:2003vg}, probing $10^8$ random points. We trade the soft trilinear parameters $A_{\lambda}, A_{\kappa}, A_{t}$ for $m_{P}, m_{A}, X_{t}$, defined by \cite{Miller20043,Ellwanger:2011sk}

\begin{eqnarray}
m_{A}^{2} & = & \frac{\lambda v_{s}}{\sin 2\beta} \left( \sqrt{2} A_{\lambda} + \kappa v_{s} \right) \label{eqma}\\
m_{P}^2 & = & - \frac{3}{\sqrt{2}} \kappa v_s A_{\kappa} \label{eqmp}\\
X_{t} & = & A_{t} - \mu_{eff}/\tan \beta
\end{eqnarray}

\noindent Here $m_{A}$ ($m_{P}$) is the diagonal component of the CP odd doublet (singlet) mass matrix and $X_{t}$ is the stop mixing parameter.

The parameters scanned are $\lambda$, $\kappa$, $m_{A}$, $m_{P}$, $\mu_{eff}$, $\tan \beta$, $M_2$, $X_{t}$ and $m_{Q_{3}}$. We fix the gaugino masses $M_{1}$ and $M_{3}$ with the unification constraints $M_{1}=\frac{1}{2} M_{2}$ and $M_{3}=3M_{2}$. We further assume $m_{Q_{3}}=m_{U_{3}}$.  All other squark and soft trilinear parameters are fixed to $1500$~GeV, and the slepton mass parameters are fixed to $200$~GeV. See Table \ref{tab:scan} for scanned parameter ranges.

\begin{table}[t]
\begin{center}
\begin{tabular}{|c|c|c|} \hline
Parameter & Range/Value & $h_{60}$ \\ \hline
 $\lambda$ & (0,0.1] & 0.03505\\
 $\kappa$ & [-0.01,0.01] & 0.006088 \\
 $m_{A}$ & [500,1500]~GeV & 1068.~GeV \\
  $m_{P}$ & [9.9,10.5]~GeV & 10.25~GeV\\
  $\mu_{eff}$ & [100,300]~GeV & 166.7~GeV \\ 
  $\tan \beta$ & [1,30] & 15.49 \\ \hline
 $M_{1}$ & $\frac{1}{2}M_{2}$ & 80.73~GeV\\
 $M_{2}$ & [100,300]~GeV & 161.5~GeV \\
 $M_{3}$ & $3M_{2}$ & 484.4~GeV\\   \hline
 $X_{t}$ & [$0.8X_{t}^{max},1.8X_{t}^{max}$] & 1378.~GeV\\
 $m_{\tilde{Q3}_{L}}$ & [350,550]~GeV & 546.9~GeV\\
 $m_{\tilde{U3}_{R}}$ & $m_{Q3}$ & 546.9~GeV \\ \hline
\end{tabular}
\caption{NMSSM parameters and their scan ranges. Additionally, $\kappa$ is constrained to satisfy $0.125 \lambda < \vert \kappa \vert < 0.225 \lambda$. 
The point $h_{60}$ ($\kappa=0.006088$ and $A_{\kappa}=-1.087$~GeV)  is taken from points surviving the scan and is described in Section \ref{sec:point}} 
\label{tab:scan}
\end{center}
\end{table}

Motivated by the PQ symmetric NMSSM, we scan small $\kappa$ and $A_{\kappa}$ or equivalently, from Equation~\ref{eqmp}, small $\kappa$ and small $m_{P}$. The lower range bound  of  $m_{P}$ (9.9~GeV) is informed by the anomalous muon magnetic moment study \cite{Gunion:2008dg}, while the upper bound (10.5~GeV) is informed by the $\eta_{b}$ mass spectrum study \cite{Domingo:2009tb}. Then $\kappa$ is scanned in the range  $-0.01 < \kappa < 0.01$ and is also required to satisfy $0.125 \lambda < \vert \kappa \vert < 0.225 \lambda$ since this requires $m_{h_1} \approx 60$~GeV within several GeV. We scan moderately small $\lambda$ in the range $0 < \lambda < 0.1$. Since the $h_{125}$ signal strength constraints are applied in the scan,  $m_{A}$ is allowed  to go into the effective MSSM decoupling limit $m_{A} \gg m_{Z}$ to accommodate the SM-like couplings of the $h_{125}$.

The neutralino and chargino masses are largely determined by $\mu_{eff},M_{1}$ and $M_{2}$ which, from naturalness considerations are bounded above in the scan by 300~GeV \cite{Papucci:2011wy}.  
At tree level, the stop masses are $m_{\tilde{t}_1,\tilde{t}_2}^{2} = m_{Q_3}^{2} + m_{t}^{2} \pm m_{t} X_{t}$ for $m_{Q_{3}}=m_{U_{3}}$ \cite{Miller20043}. Naturalness informs the $m_{Q_3}$ range since light stops are compatible with small fine tuning.

The tree level Higgs mass in the MSSM is bounded by $m_{h}^{2} < m_{Z}^{2} \cos^{2} 2\beta$, requiring a large loop correction for the $h_{125}$. In the NMSSM the upper bound on $m_{h}^{2}$ has an additional $\mathcal{O}(\lambda^{2} v^{2})$ term. The stop mixing parameter $X_t$ partly determines the one loop correction  \cite{Papucci:2011wy}:

\begin{eqnarray}
\delta m_{h}^{2} & = & \frac{3G_F}{\sqrt{2} \pi^2} m_{t}^{4} \left[ \log \left( \frac{m_{\tilde{t}}^{2}}{m_{t}^{2}} \right) + \frac{X_{t}^{2}}{m_{\tilde{t}}^{2}} \left( 1 - \frac{X_{t}^2}{12 m_{\tilde{t}}^{2}}  \right) \right]
\end{eqnarray}

\noindent where the parameter $m_{\tilde{t}}$ is defined by $m_{\tilde{t}}^{2} \equiv \frac{1}{2} ( m_{\tilde{t}_{1}}^{2}+m_{\tilde{t}_{2}}^{2} )$. The correction is strongly dependent on the top mass $m_{t}$. In the scan $m_{t}=172.5$~GeV.

To allow the large correction required by the $h_{125}$, but with small $m_{\tilde{t}}$ required by Natural SUSY, the stop mixing $X_{t}$ is allowed to contribute up to its maximal possible correction at $X_{t}^{max}=\sqrt{6}m_{\tilde{t}}$. In the scan NMSSMTools4 calculates the Higgs mass spectrum at one-loop level including external momentum for self-energies and two-loop level excluding external momentum \cite{Degrassi:2009yq,Staub:2015aea}.

\section{Surviving Points}

The suite of constraints imposed by NMSSMTools4 while scanning includes experimental results from a wide variety of sources, including:

\begin{itemize}

\item Anomalous muon magnetic moment $\Delta a_{\mu}$ measured by BNL E821 \cite{Bennett:2006fi}

\item DM relic density $\Omega_{DM} \hbar^{2}$ measured by Planck \cite{Ade:2013zuv}, direct DM exclusion by LUX \cite{Akerib:2013tjd}

\item $B$ Physics. $b \rightarrow s \gamma$, $B \rightarrow X_{s} \mu^+ \mu^-$, $B^+ \rightarrow \tau^+ \nu$,  $B_s \rightarrow \mu^+ \mu^-$, $\Upsilon(1s) \rightarrow a \gamma$, $\eta_b(1s)$  

\item Higgs. LHC $h_{125}$ , LEP $e^+ e^- \rightarrow Zh$, Tevatron/LHC $t \rightarrow b H^+$, NMSSM  searches

\item SUSY. Tevatron/LHC $\chi^+, \tilde{q}, \tilde{g}, \tilde{e}, \tilde{\mu}, \tilde{\tau}$ mass constraints, $\tilde{t} \rightarrow b \ell \tilde{\nu}, \chi^0 c$ and $\tilde{b} \rightarrow \chi^0 b$

\end{itemize}

\noindent Loose constraints imposed during the scan require $m_{h_{2}} \approx 125$~GeV within 3~GeV and impose an upper bound on each $h_{125}$ signal strength $\chi^2$, calculated as in \cite{Belanger:2013xza}. Of the $10^{8}$ points scanned, $42$ survive the constraints imposed during the scan.

Constraints are tightened after the scan. The low mass Higgs sector must satisfy: 

\begin{center}
$9.9  <  m_{a_1}  < 10.5$~GeV

$50 < m_{h_1} < 70$~GeV 

$122 < m_{h_2} < 128 $~GeV
\end{center}

\noindent Finally, the sum of $h_{125}$ signal strength $\chi^2$ are required to satisfy $\sum_{i} \chi_{i}^{2} < 13$. Of the $42$ points surviving the scan constraints, $15$ points survive these final constraints. 

In order to demonstrate the naturalness of the surviving points, we examine the fine tuning metric $F_{max} \equiv max_{a \in A} \left( \frac{\partial ( \log m_{Z}^{2}}{\partial (\log a^2)} \right)$ calculated by NMSSMTools4. This metric yields the largest fine tuning over fundamental parameter set $A$. Surviving points have small fine tuning, $ 5 < F_{max} < 10$, light stops $300 < m_{\tilde{t}_{1}}<400$~GeV and light gluinos $500 < m_{\tilde{g}} < 650$~GeV. While agreement is not universal on which $F_{max}$ values characterize low fine tuning \cite{Schuster:2005py}, studies have considered $F_{max}$ of order $\mathcal{O}(10^2)$ to be typical  for the NMSSM \cite{Binjonaid:2014oga} and  $\mathcal{O}(10^1)$ to be low fine tuning \cite{Dermisek:2007yt,Dermisek:2006wr,Dermisek:2005ar,Hall:2011aa}. A recent study seeking to establish naturalness as objective, model-independent and predictive concludes that a SUSY model with $F_{max}<30$ is natural, while one with $F_{max}<10$ is stringently natural \cite{Baer:2015rja}.

\section{Benchmark $\mathbf{h_{60}}$ \label{sec:point}}

\begin{figure}[t]
\begin{center}
\includegraphics[width=0.49\textwidth]{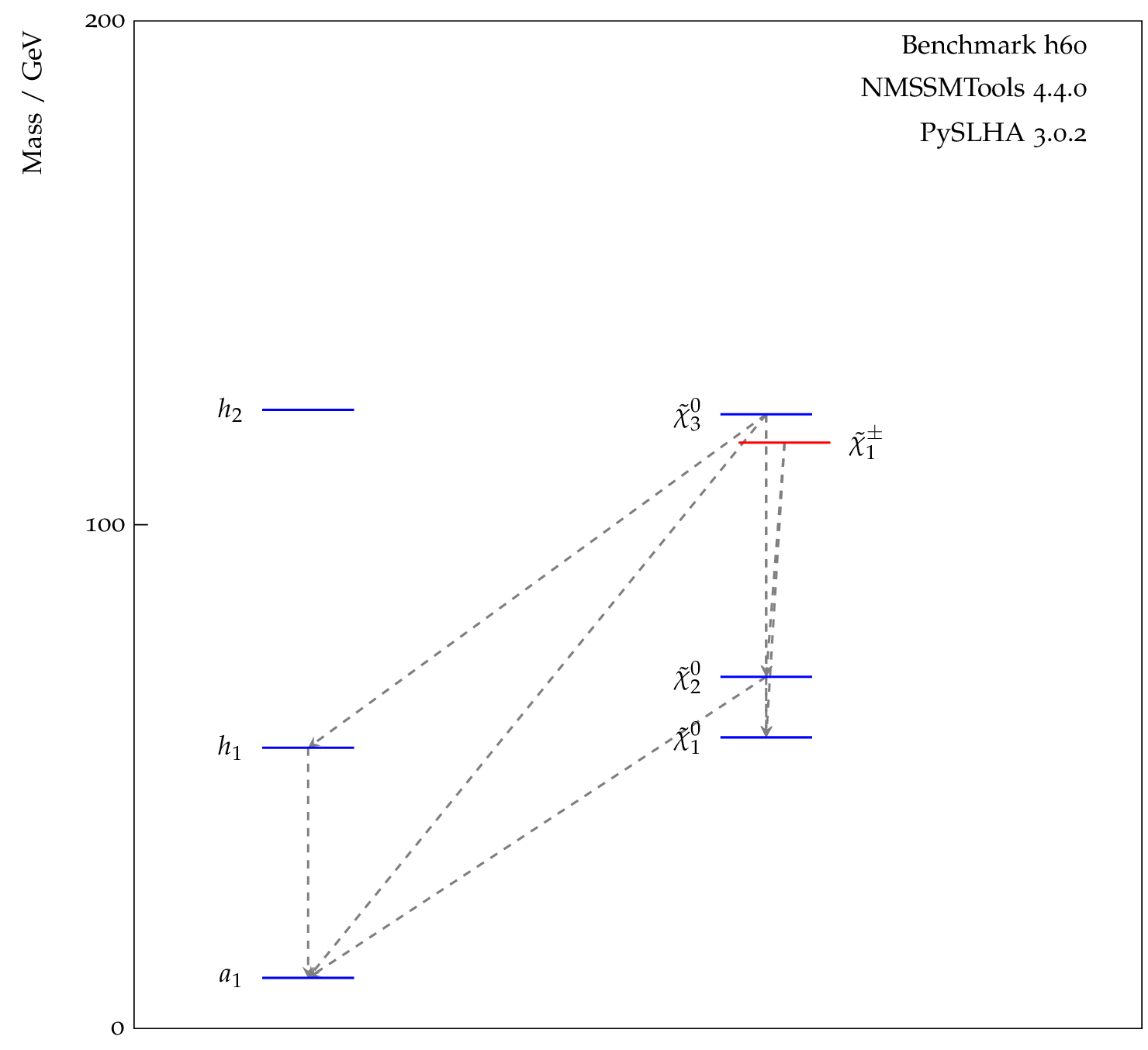}
\caption{Masses and decays for the lighter $h_{60}$ benchmark spectrum. Only decays with branching ratios larger than 5\% are shown. Note that the $h_{2}$ is mostly decoupled.}
\label{fig:pyslha}
\end{center}
\end{figure}

\begin{table}[t]
\begin{center}
\begin{tabular}{|c|c|c|c|c|c|} \hline
Component & $a_1$ & $h_1$ & $\chi_1$ & $\chi_2$ & $\chi_3$ \\ \hline
$P_{11}/S_{11}/N_{i1}$ & -0.002 & -0.006 & 0.151 & 0.882 & 0.401 \\
$P_{12}/S_{12}/N_{i2}$ & 0.000 & -0.130 & -0.054 & -0.153 & 0.679 \\
$P_{13}/S_{13}/N_{i5}$ & 1.000 & 0.992 & 0.981 & -0.189 & 0.045 \\ \hline 
\end{tabular}
\caption{Doublet and singlet components of the $a_1$, $h_1$ and gaugino and singlino components of the $\chi_1,\chi_2$,$\chi_3$ in $h_{60}$.}
\label{tab:components}
\end{center}
\end{table}

It has been noted that points surviving the scan represent a Natural NMSSM with slightly broken PQ symmetry. They also exhibit a light pseudoscalar Higgs with $m_{a_1} \approx 10$~GeV, a light scalar Higgs with $m_{h_1} \approx 60$~GeV, a singlino LSP DM candidate with $m_{\chi_{1}} \approx 60$~GeV annihilating via $\chi_1 \chi_1 \rightarrow b\bar{b}$, and a light stop with $m_{\tilde{t}_{1}} \approx 350$~GeV.

The benchmark point $h_{60}$ satisfies the threshold criterion $m_{\chi_{3}}>m_{h_1} +m_{\chi_1}$ with the largest branching ratio for $\chi_{3} \rightarrow \chi_1 h_1$ of all surviving points in the scan. The lowest branching ratio for this decay in the surviving points which reach threshold is 65\%, while the highest is 80\%. This ensures production of $a_1$ from $h_1 \rightarrow a_1 a_1$ in stop pair events with $\tilde{t}_{1} \rightarrow \chi_{2}^{+} b \rightarrow \chi_{3} Wb$ or $\tilde{t}_{1} \rightarrow \chi_{3} t$. See the last column of Table~\ref{tab:scan} for the numerical values of the parameters which define $h_{60}$.   See Figure~\ref{fig:pyslha}, generated with PySLHA \cite{Buckley:2013jua} using the SUSY Les Houches Accord (SLHA) \cite{Skands:2003cj,Allanach:2008qq}  file produced by NMSSMTools4, for the mass spectrum and decays in $h_{60}$. 

For components $P_{1j},S_{1j},N_{ij}$ of the $a_1,h_1,\chi_1,\chi_2$ and $\chi_3$ in $h_{60}$ see Table \ref{tab:components}.  The LSP $\chi_{1}$ is dominantly singlino, while the $a_1$ and $h_1$ are dominantly singlet. The $\chi_{1}$ mixing with doublinos and gauginos is small, as is the mixing of the $a_1$ and $h_1$ with the doublets. The phenomenology of a singlino LSP at the LHC has been considered in \cite{Ellwanger:1997jj,Gunion:2005rw,Cao:2013mqa,Han:2014nba,Ellwanger:2014hia,Kim:2014noa,Han:2015zba,Draper:2010ew,Casolino:2015cza}. The phenomenology of light stops in the NMSSM, and how they avoid exclusion at the LHC,  has been recently considered in \cite{Beuria:2015mta}. 

\begin{table}[t]
\begin{center}
\begin{tabular}{|c|c|c|c|c|} \hline
 &  Range & Mass [GeV]  & BR1 (\%) & BR2 (\%) \\ \hline
$a_1$ & [9.9,10.4] & 10.0 &  $\tau^+ \tau^-$ (81)  &  $gg$ (16)\\
$h_1$ & [53,59] & 55.7 & $a_1 a_1$ (72) & $b \bar{b}$ (23) \\
$h_2$ & [122,123] & 122.8 & $b \bar{b}$ (65) & $W W^{\star}$ (17)  \\ \hline
$\chi_{1}$ & [54,60] & 57.8  & - & - \\
$\chi_{2}$ & [57,76] & 69.8  & $\chi_{1} a_{1}$(75) & $\chi_{1} Z^{\star}$ (25)  \\
$\chi_{3}$ & [107,136] & 121.9  & $\chi_{1} h_{1}$ (80)  & $\chi_{2} Z^{\star}$ (10) \\
$\chi_{4}$ & [166,209] & 179.5  & $\chi_{2} Z$ (80)& $\chi_{1} Z$ (14)\\
$\chi_{5}$ & [225,254] & 236.6  & $\chi^{+}_{1} W$ (60) & $\tilde{\nu} \nu,\tilde{\ell} \ell$ (38) \\ \hline
$\chi_{1}^{+}$ & [104,132] & 116.3  & $\chi_{2} W^{\star}$ (78) & $\chi_{1} W^{\star}$ (22)  \\
$\chi_{2}^{+}$ & [225,255] & 237.1 & $\ell \tilde{\nu}, \tilde{\ell} \nu$ (40) & $\chi_{3} W$ (38) \\ \hline
$\tilde{t}_{1}$ & [313,391]& 335.6  & $\chi^{+}_{2} b$ (75) & $\chi_{3} t$ (15) \\ \hline
%$\tilde{b}_{1}$ & [486,545] & 544.7  & $\tilde{t}_{1} W$ (41) & $\chi_{2}^{+} t$ (31) \\
%$\tilde{g}$ &  [505,637] & 611.2 & $\tilde{t}_{1} t$ (94) & $\tilde{b}_{1} b$ (6)  \\ \hline 
\end{tabular}
\caption{Masses and the two dominant decays of the lighter part of the $h_{60}$ benchmark point spectrum obtained by NMSSMTools4. The column titled ``Range'' is the mass range of the points surviving all scan constraints.}
\label{tab:spectrum}
\end{center}
\end{table}

For the numerical values of the masses and dominant branching ratios of the low mass spectrum of the $h_{60}$ benchmark, see Table~\ref{tab:spectrum}. The $a_1$ and $h_1$ of the benchmark avoids the LHC search exclusion for straightforward reasons. Both ATLAS and CMS have searched for gluon fusion $gg \rightarrow a \rightarrow \mu^+ \mu^-$  but critically omit the $\Upsilon$ region and therefore cannot exclude $m_{a_1} \approx 10$~GeV \cite{ATLAS-CONF-2011-020,CMS-PAS-HIG-12-004}. ATLAS has searched for gluon fusion $gg \rightarrow h \rightarrow aa$ for $2m_{\tau} < m_a < 2m_{B}$  but does not report limits for $m_{h}<100$~GeV \cite{Aad:2015oqa}. CMS has searched for the same channel, but only reports limits  for $m_{h}>90$~GeV with $m_{a}<2m_{\tau}$ \cite{Khachatryan:2015wka} or for the $h_{125}$ with $4 < m_{a}<8$~GeV \cite{Khachatryan:2015nba}. More decisively, the gluon fusion cross sections for $a_1$ and $h_1$ production in the benchmark are greatly reduced relative to the $h_{125}$.

In the neutralino and chargino sector, both ATLAS and CMS have studied $\chi_{2} \chi^{+}_{1}$ production \cite{Khachatryan:2014qwa,Khachatryan:2014mma,Aad:2015eda}. For example, the searches which assume decays to sleptons or to bosons also assume that $m_{\chi^{+}_{1}}=m_{\chi_{2}}$, motivated by  models with a bino-like $\chi_{1}$ and wino-like $\chi_{2}$ and $\chi^{+}_{1}$. But in $h_{60}$ the $\chi_{1}$ is singlino, and manifestly $m_{\chi^{+}_{1}} \neq m_{\chi_{2}}$. 
The $\chi_{2} \chi^{+}_{1}$ searches which assume dominant decays to sleptons cannot exclude $h_{60}$ where $m_{\tilde{\ell}}>m_{\chi^{+}_{1}},m_{\chi_{2}}$. Such searches might be sensitive to $\chi_{5} \chi^{+}_{2}$ events, but here the cross section is reduced and the final states are more complex. Of the searches which assume dominant decays to bosons, only the $W \chi_{1}Z \chi_{1}$ final state case applies. In this case both $W$ and $Z$ are very far off mass shell in $h_{60}$, in which case it is unlikely that the exclusion can apply.

In the stop sector, both ATLAS and CMS report exclusion. For a summary of the ATLAS results, see \cite{Aad:2015pfx,Aad:2015iea}. For a bibliography of CMS results see \cite{CMS-DP-2015-035}.  No exclusion is given for the NMSSM, however, and exclusion for simplified models cannot be easily interpreted in the NMSSM context. For example, the analyses which assume $\tilde{t} \rightarrow t \chi_{1}$  with 100\% branching ratio cannot exclude $h_{60}$, for which this branching ratio is  $\mathcal{O}(10^{-3})$. $h_{60}$ does contain $\tilde{t} \rightarrow t \chi_{3}$ with a branching ratio $\mathcal{O}(10^{-1})$, but the subsequent $\chi_{3}$ decay produces a much more complex final state with less missing energy than assumed by the searches. 
The stop pair searches which assume $\tilde{t} \rightarrow b \chi^{+}_{1}$ with 100\% branching ratio assume very specific cases of mass relationships between stops, neutralino and charginos which do not hold in $h_{60}$. Moreover they assume $\chi^{+}_{1} \rightarrow W \chi_{1}$. But the dominant branching in $h_{60}$ is $\chi^{+}_{1} \rightarrow W^{\star} \chi_{2} \rightarrow W^{\star} Z^{\star} \chi_{1}$, producing less missing energy, and two gauge bosons which are very far off mass shell in comparison to the search assumptions.

\begin{table}[t]
\begin{center}
\begin{tabular}{|c|c|c|c|c|} \hline
Sample & $r_{max}$ & Analysis & $\int dt \mathcal{L}$ [fb$^{-1}$] \\ \hline
$\chi \bar{\chi}$ & 0.6 & atlas\_conf\_2013\_035 &  20.7\\
$\tilde{t_1} \tilde{\bar{t_1}}$ & 0.5 & atlas\_conf\_2013\_061 & 20.1 \\
$\tilde{g} \tilde{g}$ & 12.5 & atlas\_conf\_2013\_061 & 20.1 \\ \hline
$\chi \bar{\chi}$ & 0.1 & cms\_1303\_2985 & 11.7\\
$\tilde{t_1} \tilde{\bar{t_1}}$ & 0.6 & cms\_1502\_06031 & 19.5 \\
$\tilde{g} \tilde{g}$ & 1.2 & cms\_1303\_2985 & 11.7\\ \hline
\end{tabular}
\caption {Maximum exclusion $r_{max}$ determined by Checkmate1 of the analyses most sensitive to $h_{60)}$ of all the validated ATLAS and CMS Run 1 analyses. Note that neither stop nor chargino/neutralino pair production is ruled out for $h_{60}$.}
\label{tab:checkmate}
\end{center}
\end{table}

In order to evaluate quantitatively the $h_{60}$ exclusion at LHC Run 1, we run all 16 (3) presently validated ATLAS (CMS) analyses in Checkmate1.2.2 \cite{Drees:2013wra,deFavereau:2013fsa,Cacciari:2011ma,Cacciari:2005hq,Cacciari:2008gp,Read:2002hq} on generated $h_{60}$ events. Event simulation of the gluino, stop and chargino/neutralino pair production is carried out with Pythia8.205 \cite{Sjostrand:2007gs,Sjostrand:2006za}. The SLHA  file produced by NMSSMTools4 for $h_{60}$ is used with Pythia8, which features a dedicated NMSSM model with functionality for SLHA input. 

See Table~\ref{tab:checkmate} for the exclusion $r_{max}$, the ratio of the 95\% confidence level lower limit on the $h_{60}$ signal presence to the measured 95\% confidence level limit, of the analyses with maximum sensitivity to $h_{60}$ chargino/neutralino, stop and gluino pair production. Only for gluino pair production is $r_{max}>1$, indicating that both ATLAS and CMS have ruled out a gluino with $m_{\tilde{g}}\approx 611$~GeV in $h_{60}$ but neither has ruled out the stop and chargino/neutralino sectors of $h_{60}$. However, since $m_{\tilde{g}}$ is determined by the gaugino mass $M_{3}$, which can be easily increased without otherwise impacting the lower energy $h_{60}$ phenomenology, we simply assume $m_{\tilde{g}} \approx 855$~GeV or greater since this reduces the gluino pair production cross section by a factor of 13 relative to $h_{60}$.

Note that $r_{max}=0.6$ for cms\_1502\_06031, which exhibits a $3\sigma$ excess in the low dilepton mass region \cite{Khachatryan:2015lwa}. If the $h_{60}$ stop mass is reduced such that the stop pair production cross section is enhanced by a factor of 1.5, then this analysis becomes sensitive to $h_{60}$ with the reduced $m_{\tilde{t}} \approx 315$~GeV.

 \section{Collider Signature}

Since the stop is relatively light in $h_{60}$, the cross section for pair production is large and makes cascade production of the $a_1$ and $h_1$ accessible at the LHC. Gluon fusion production of $a_1$ and $h_1$ is less promising. The reduced $tth_1$ ($tta_1$) coupling, which appears in the gluon fusion top loop, is of order $\mathcal{O}(10^{-1})$  ($\mathcal{O}(10^{-5})$) relative to the SM $ttH_{SM}$ coupling for a SM Higgs boson of the same mass. 

From Table~\ref{tab:spectrum} the decays $\tilde{t}_{1} \rightarrow \chi_{2}^{+}b $, $\chi_{2}^{+} \rightarrow \chi_{3} W$, $\chi_{3} \rightarrow \chi_{1} h_{1}$, and $h_1 \rightarrow 2a_1$ proceed with branching ratios of 75\%, 38\%, 80\%, and 72\% respectively, while $\tilde{t}_{1} \rightarrow \chi_{3} t$ proceeds with branching ratio of 15\%. This makes stop pair production with $\tilde{t}_{1} \rightarrow \chi_{2}^{+} b \rightarrow \chi_{3} W b$ or $\tilde{t}_{1} \rightarrow \chi_{3} t$ and $\chi_{3} \rightarrow \chi_{1} h_{1} \rightarrow \chi_{1}2a_1$  promising channels for discovery if the $a_{1}$ can be successfully reconstructed, for example in the relatively rare but very clean $a_{1} \rightarrow \mu^+ \mu^-$ channel. In $h_{60}$ this decay proceeds with a branching ratio of 0.3\%. These remarks also apply to stop pairs produced in gluino pair production with $\tilde{g} \rightarrow \tilde{t}_{1} t$.

In $h_{60}$ stop pair production, the cascade dominantly contains two top quarks. In gluino pair production it dominantly contains four top quarks. These are strong handles on any potential background. Some top pair $t\bar{t}$ background may be irreducible, but other backgrounds should be negligible.

We now describe a targeted study of the sensitivity to $h_{60}$ at the LHC.  Signal events are generated with Pythia8 as described in Section~\ref{sec:point}. Background $t\bar{t}$ events are also generated in Pythia8.  Fast detector simulation is performed with Delphe3.2.0 \cite{deFavereau:2013fsa}. The Delphes3 detector card for CMS is modified to reproduce the tight electron, tight muon and $b$ tag efficiencies reported by CMS \cite{CMS-DP-2013-003,CMS-PAS-BTV-13-001,Chatrchyan:1456510}.  The signal selection seeks the decay $a_1 \rightarrow \mu^+ \mu^-$ in gluino and stop pair events and employs a standard selection for semileptonic top pair events, together with a selection for $a_{1} \rightarrow \mu^+ \mu^-$, in which one top quark decays via $t \rightarrow bW \rightarrow b\ell \nu$ and the other via $t \rightarrow bW \rightarrow bq q^{\prime}$. The requirements for the Run 1 analysis are these:

\begin{itemize}

\item exactly one tight electron with $E_{T}>25$~GeV and no isolation requirement

\item missing transverse energy $E_{T}^{miss}>85$~GeV

\item four or more jets with $E_{T}>20$~GeV, at least two of which are $b$-tagged

\item two or more tight muons with $p_{T}>2$~GeV, no isolation requirement and $d_{0}/\sigma_{d_0}<5$

\item zero net charge and  $9.7 < m_{\mu^+ \mu^-} < 10.3$~GeV in the leading and subleading muons

\end{itemize}

\noindent The $a_1$ candidate is then reconstructed from the leading and subleading muons. The muon azimuthal impact parameter significance requirement $d_{0}/\sigma_{d_0}<5$ ensures that the muons are consistent with prompt production. 

For the Run 2 analysis, we assume $\sqrt{s}=14$~TeV and $\int dt \mathcal{L}=300$~fb$^{-1}$. We use the Delphes3 simulation with mean pileup 50. The selection is identical to the Run 1 analysis except that the electron, jet, and muon thresholds are raised to 30~GeV, 30~GeV and 4~GeV, respectively.

\begin{table*}[t]
\begin{center}
\begin{tabular}{|c|ccc|ccc|ccc|} \hline
 Process&  \multicolumn{3}{|c|}{Run 1: $\mathcal{L}=5$~fb$^{-1}$} & \multicolumn{3}{|c|}{Run 1: $\mathcal{L}=20$~fb$^{-1}$} & \multicolumn{3}{|c|}{Run 2: $\mathcal{L}=300$~fb$^{-1}$} \\
& $\sigma_{7T}$ [pb] & $N_{n}$ & $N_{p}$ & $\sigma_{8T}$ [pb] & $N_{n}$ & $N_{p}$ & $\sigma_{14T}$ [pb] & $N_{n}$ & $N_{p}$\\ \hline
SM $t\bar{t}$ & 173.6 & 1.6 & 0  & 247.7 & 8.5 & 0 & 966.0 & 286.0 & 0 \\ 
%$\tilde{g}\tilde{g}$ & 0.6 & 15.4(8.9) & 8.6(10.2) & 1.2 & 108.0 & 65.4 & 10.9 & 11599.6 & 7551.5\\
$h_{60}$ $\tilde{t}_{1} \tilde{\bar{t}}_{1}$ & 0.7 & 2.6(0.8) & 2.2(2.5) & 1.1 & 20.1 & 19.2  & 5.9 & 892.5 & 647.9\\ \hline
$S/\sqrt{B}$ & \multicolumn{3}{|c|}{1.0(1.6)} & \multicolumn{3}{|c|}{3.6} & \multicolumn{3}{|c|}{18.9}  \\ \hline
\end{tabular}
\caption{NLO cross sections for $\tilde{t}_{1} \tilde{\bar{t}}_{1}$ and $t\bar{t}$ production at the LHC in Runs 1 ($\sqrt{s}=7,8$~TeV) and 2 ($\sqrt{s}=14$~TeV). We omit the $\tilde{g}\tilde{g}$ yields. Also shown are the expected yields for peaking events ($N_{p}$), yields for nonpeaking events ($N_{n}$) and signal significances after the full signal selection described in the text. For $\sqrt{s}=7$~TeV, we show in parentheses yields and significance for a variation of $h_{60}$ in which the $\chi_2$ decays outside of the effective tracking volume.}\label{tab:lhc}
\end{center}
\end{table*}

\begin{figure}[b]
\begin{center}
\includegraphics[width=0.48\textwidth]{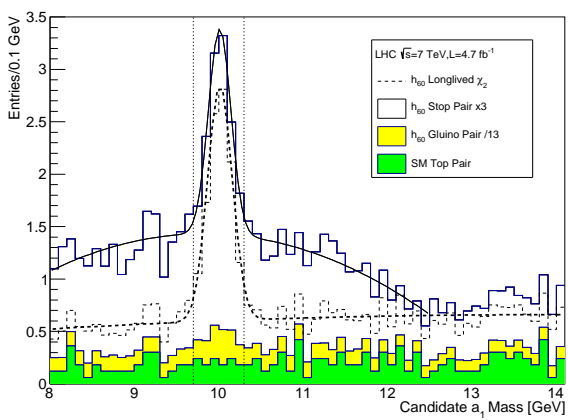}
s\caption{Reconstructed candidate $a_{1} \rightarrow \mu^+ \mu^-$ mass distribution after full signal selection assuming $h_{60}$  ($\times 3$ for $\tilde{t_1} \tilde{\bar{t_1}}$ and $/13$ for $\tilde{g}\tilde{g}$) at the LHC for $\sqrt{s}=7$~TeV and $\int dt \mathcal{L}=4.7$~fb$^{-1}$. Also shown (dashed) is a variation of $h_{60}$ in which the $\chi_{2}$ decays outside of the effective tracking volume. The fits employ a Gaussian signal model and a polynomial background model.} 
\label{fig:lhc}
\end{center}
\end{figure} 

After full signal selection, the SM top background is nearly negligible. Multiple jet events produced by QCD have not been simulated, but with the nominal selection this background is expected to be very small. In data, the nonpeaking $h_{60}$ events in the candidate $a_1$ distribution can be mistaken for QCD multijets events, however, so these are considered background in the significance calculation. In the SM top background, the candidate $a_{1}$ muons originate from $\tau$ lepton, $D$ meson or $B$ meson decays. In the nonpeaking $h_{60}$ events, they originate either from SM $\tau$, $B$ or $D$ decay or from NMSSM $a_{1} \rightarrow \tau_{\mu} \tau$, $\chi_{2} \rightarrow \chi_{1} \mu^{+} \mu^{-}$, $\chi^{+} \rightarrow \chi \mu \nu$, or $\chi^+ \rightarrow \mu \tilde{\nu}$. 

The proportion of peaking to nonpeaking signal events is sensitive to the details of $h_{60}$. For example, if the slepton masses are raised above the threshold for decay from $\chi_2^{+}$, then the peaking signal is enhanced and the nonpeaking signal is reduced. Similarly, if the branching ratio for $\chi_2 \rightarrow \chi_1 a_1$ is raised at the expense of $\chi_2 \rightarrow \chi_1 Z^{\star}$, the peaking signal is enhanced and the nonpeaking signal is reduced. Finally, if the $\chi_2$ width is sufficiently small, its decay vertices may lie outside the effective tracking volume, making nonpeaking background from $\chi_2$ effectively invisible. 

Pythia8 is a leading order generator, but next to leading order cross sections obtained  by the LHC SUSY Working Group \cite{Kramer:2012bx,Borschensky:2014cia} are used to normalize the event yields.  See Figure~\ref{fig:lhc} for the reconstructed $a_1$ mass distribution after full signal selection, where the distribution for a variation of $h_{60}$ in which the $\chi_2$ decays outside of the tracking volume is also shown. See Table~\ref{tab:lhc} for expected peaking and nonpeaking event yields and signal significances after full selection for Runs 1 and 2 at the LHC. A targeted $h_{60}$ signal selection yields sensitivity even at the LHC Run 1.

The advantages of the International Linear Collider (ILC) for studying low mass NMSSM Higgs bosons has been noted in \cite{Miller20043}. At the ILC, the standard Higgstrahlung production channel $e^{+} e^{-} \rightarrow Z h_{1}$ is suppressed in the NMSSM due to the measured SM-like $h_{125} \rightarrow ZZ^{\star}$ signal strength and the NMSSM coupling sum rule $\sum_{1=1}^{3} \xi_{ZZh_{i}}^{2} =1$ \cite{Ellwanger:2009dp}. Instead, in $h_{60}$ we note the possibility of resonant production $e^{+}e^{-} \rightarrow a_1 h_1$, with cross section of several hundred picobarns at $\sqrt{s}=m_{Z}$. For $\sqrt{s}=500$~GeV, pair production of all neutralinos and all charginos is accessible with cross sections nearing a picobarn, as well as $a_1 h_1$ and $Zh_2$ production cross sections of about a hundred femtobarns.

In $h_{60}$ the $Za_1 h_1$ coupling is small enough to have evaded LEP searches \cite{Barate:2003sz,Alexander:1996ai} but large enough to be produced copiously at the ILC running on the $Z$ pole. 
The ILC sensitivity to $h_{60}$ in operating scenarios described in \cite{Barklow:2015tja} defined by beam polarization, luminosity and $\sqrt{s}$ and will be evaluated in a forthcoming companion study.

\section{Conclusion}

We have reviewed the motivation for a natural NMSSM with a slightly broken PQ symmetry and a lightly coupled singlet featuring a light singlet pseudoscalar $a_1$, light singlet scalar $h_1$ and a light singlino LSP $\chi_1$ DM candidate annihilating via $\chi_1 \chi_1 \rightarrow b\bar{b}$. 

A random parameter space scan is performed subject to a full suite of experimental constraints, including the anomalous muon magnetic moment, the DM relic density and collider searches. Surviving points are characterized by low fine tuning, and abundant pseudoscalar $a_1$ production identifies the benchmark point $h_{60}$. In addition this benchmark features light stops and light higgsinos, all characteristic of Natural SUSY.

The benchmark avoids the current LHC exclusion limits. For the $a_1$ and $h_1$, this is due to the reduced gluon fusion cross sections. For other SUSY searches, this is primarily due to the search assumption that stops, neutralinos and charginos will decay directly to the LSP $\chi_1$ with no intermediate SUSY particles in the decay chain. But in the $h_{60}$ benchmark the $\chi_1$ is singlino and couples weakly to the rest of SUSY. Thus due to the light mass spectrum the decay chains can contain many intermediate SUSY particles, making the final states more complex with less missing energy than in the simplified search scenarios. 

Finally, we report that the potentially fruitful discovery channels at the LHC for the benchmark considered are stop and gluino pair production with either $\tilde{t}_{1} \rightarrow \chi_{2}^{+} b \rightarrow \chi_{3} Wb$ or $\tilde{t}_{1} \rightarrow \chi_{3} t$ and $\chi_{3} \rightarrow \chi_{1} h_{1} \rightarrow \chi_{1} a_1 a_1$. We conclude with a fast simulation study that with a targeted signal selection the LHC may already be sensitive to $h_{60}$ in Run 1. We have also pointed out the possibility to observe at the ILC resonant $e^+ e^- \rightarrow a_1 h_1$ at $\sqrt{s}=m_{Z}$ and pair production of all neutralinos and charginos at $\sqrt{s}=500$~GeV. 

\begin{acknowledgement}
\begin{center}\textbf{Acknowledgements}\end{center}
The author thanks M. Muhlleitner for dialog on the Higgs mass calculation in NMSSMTools4, Tao Liu for feedback on fine tuning and the PQ symmetric NMSSM, and the Alder Institute for High Energy Physics for financial support. 
\end{acknowledgement}

\bibliography{paper}

\end{document}